\documentclass[twocolumn]{IEEEtran}

\usepackage{amsmath}
\usepackage{amssymb}
\usepackage{amsfonts}
\usepackage{mathrsfs}
\usepackage{epsfig}
\usepackage{bm}
\usepackage{color}
\usepackage{cite}

\begin{document}

\title{One-to-one full scale simulations of laser wakefield acceleration using QuickPIC}

\author{J. Vieira, F. Fi\'uza, R.A. Fonseca, L.O. Silva \thanks{J. Vieira, F. Fi\'uza, R.A. Fonseca, L.O. Silva, are with GoLP/Instituto de Plasmas e Fus\~ao Nuclear, Instituto Superior T\'ecnico, 1049-001 Lisboa, Portugal (e-mail: jorge.vieira@ist.utl.pt, luis.silva@ist.utl.pt)} \thanks{R. A. Fonseca is also with Departamento de Ci\^encias e Tecnologias da Informa\c{c}\~ao, Instituto Superior de Ci\^encias do Trabalho e da Empresa, 1649-026 Lisboa, Portugal}, C. Huang, W. Lu, M. Tzoufras, F.S. Tsung, V. Decyk, W.B. Mori, \thanks{C. Huang, W. Lu, M. Tzoufras, F.S. Tsung, V. Decyk, W.B. Mori are with University of California, Los Angeles, CA 90095}J. Cooley, T. Antonsen Jr.\thanks{J. Cooley, T. Antonsen Jr. are with University of Maryland, USA}}

\maketitle

\begin{abstract}
We use the quasi-static particle-in-cell code QuickPIC to perform full-scale, one-to-one LWFA numerical experiments, with parameters that closely follow current experimental conditions. The propagation of state-of-the-art laser pulses in both preformed and uniform plasma channels is examined. We show that the presence of the channel is important whenever the laser self-modulations do not dominate the propagation. We examine the acceleration of an externally injected electron beam in the wake generated by $\sim10~\mathrm{J}$ laser pulses, showing that by using ten-centimeter-scale plasma channels it is possible to accelerate electrons to more than $4~\mathrm{GeV}$. A comparison between QuickPIC and 2D OSIRIS is provided. Good qualitative agreement between the two codes is found, but the 2D full PIC simulations fail to predict the correct laser and wakefield amplitudes. 
\end{abstract}

\begin{keywords}
Simulation, Lasers, Plasma waves, Electron accelerators
\end{keywords}

\section{Introduction}\label{sec:introduction}

The advent of high intensity ($I>10^{18}\mathrm{W/cm^2}$), tightly focused ($W_0\approx10 \mu\mathrm{m}$) and ultra-short ($\tau_0\geq30\mathrm{fs}$) laser pulses lead to the production of quasi-mono-energetic electron bunches via Laser Wake Field Acceleration (LWFA) \cite{bib:mangles,bib:geddes,bib:faure,bib:leemans,bib:leemans2}. These lasers are used in state-of-the-art experiments, leading to the self-injection of plasma electrons once the 'bubble' or 'blowout' regime~\cite{bib:puckov,bib:lu,bib:mori} is reached. Ideally in this regime, the transverse laser ponderomotive force is strong enough to expel nearly all electrons away from the laser axis. As electrons return to the axis they cross and form a thin electron layer, leaving behind an ion column. The spherical shape of the wake has ideal properties for electron acceleration, with linear accelerating and focusing forces \cite{bib:lu}, which provides high-energy and high-quality electron acceleration.

The physics associated with the blowout regime is complex and can only be fully explored using numerical simulations, which constitute an insightful tool to examine the mechanisms associated with electron acceleration in this regime. The increased interaction length required for higher energy gains poses, however, a natural difficulty to full PIC simulations, which require more than $10^5$ CPU hours for the modeling of centimeter-scale plasmas; therefore systematic studies, modeling and planning of new LWFA experiments with full PIC simulations, become increasingly difficult. One way to overcome this difficulty is the use of reduced PIC codes, such as QuickPIC~\cite{bib:chengkun} which allow for computational time savings of more than three orders of magnitude in comparison to full PIC codes. In this paper, we will then use QuickPIC to perform one-to-one, full scale LWFA simulations, both in uniform plasmas and in plasma channels, with and without externally injected electron beams.

This paper is structured as follows. In section \ref{sec:physical} we briefly present QuickPIC physical model. In Section \ref{sec:onetoone} we examine $\sim 2~\mathrm{J}$ laser evolution, and the corresponding plasma response, using physical parameters that closely follow Refs.~\cite{bib:leemans,bib:leemans2}. In addition we present simulation results with high energy laser pulses, over longer propagation distances, towards stable multi-GeV electron acceleration. In Section \ref{sec:conclusions}, a critical discussion of the results of Section \ref{sec:onetoone} is given, and we compare both the laser evolution and driven wakefield between QuickPIC with full 2D OSIRIS~\cite{bib:fonseca} simulations. Finally, we state the conclusions.

\section{QuickPIC physical model}\label{sec:physical}

For the sake of completeness we outline here the key assumptions of the QuickPIC physical model \cite{bib:chengkun}. 

QuickPIC works under the quasi-static approximation (QSA)~\cite{bib:sprangle} which is valid whenever the typical evolution time/length for the driver is much larger than the typical time/length for the plasma response. The typical time for the evolution of a laser driver in the QSA is generally determined by the Rayleigh length, $Z_r$, while the plasma response is determined by the plasma wavelength, $\lambda_p$. Thus, the QSA is always satisfied for the typical experimental parameters required for LWFA. The QSA fails, however, to model trapped particles, precluding the physics that is present in the self-injection. Consequently, the dynamics of electrons in the blowout regime can only be examined through the evolution of an externally injected electron beam. 

QuickPIC solves the Maxwell's equations, written in the Lorentz gauge under the QSA in the co-moving frame $(x,y,\xi=z-ct,s=z)$ given by:

\begin{equation}
\label{eqn:max1}
-\nabla_{\perp}^2 \phi = 4 \pi \rho,
\end{equation}    

\begin{equation}
\label{eqn:max2}
-\nabla_{\perp}^2 \mathbf{A} = 4 \pi \mathbf{j},
\end{equation}    
where $\phi$ and $\mathbf{A}$ represent the plasma scalar and vector potentials, $\rho$ and $\mathbf{j}$ represent the charge and current distributions and $\nabla_{\perp}^2 \equiv \partial^2/\partial^2 x + \partial^2/\partial^2 y $. The charge distribution is initialized as:

\begin{equation}
\label{eqn:channel}
n(r)= n_0 \left(1+\frac{\Delta n}{n_0} \frac{r^2}{W_0^2}\right), \text{if}~r<r_{\mathrm{c}},
\end{equation}
falling linearly for $r>r_{\mathrm{c}}$. The background plasma density is denoted by $n_0$, the depth of the channel by $\Delta n$, the distance to the axis by $r=\sqrt{x^2+y^2}$ and $r_{\mathrm{c}}$ is the radius of the channel.

Eqs.~(\ref{eqn:max1}) and (\ref{eqn:max2}) reveal that the plasma response is fully determined by the transverse radial charge and current distributions. Therefore, the simulation domain is divided into transverse slices, perpendicular to the laser propagation direction, where the plasma response is determined for a given laser profile. After solving the plasma response for the whole simulation box, the laser is advanced according to the ponderomotive guiding center approximation:
\begin{equation}
\label{eqn:envelope}
\left[2 \frac{\partial }{\partial s} \left(-i k_0 - \frac{\partial }{\partial \xi}\right) - \nabla_{\perp}^2\right] \mathbf{A}^{\mathrm{laser}}= \frac{4 \pi}{c} \mathbf{j},
\end{equation}
where $\mathbf{A}^{\mathrm{laser}}$ is the laser vector potential, defined according to:
\begin{equation}
\label{eqn:laser}
a = a_0 \left(10 \tau^3-15 \tau^4+6 \tau^5\right) \exp{\left[-r^2/W_0^2\right]},
\end{equation}
where $a\equiv q A^{\mathrm{laser}}/(m_e c^2)$, $\tau=s/\tau_0$, $\tau_0$ is the laser duration at full width half maximum on the fields,  and $W_0$ is the spot of the laser.

Even though Eq.~(\ref{eqn:envelope}) evolves the envelope of the laser, the longitudinal grid size $\Delta \xi$ should be chosen as small as possible, such that it is guaranteed that the frequency shifts associated with the envelope modulations are captured correctly. Since the discretization of Eq.~(\ref{eqn:envelope}) implies a stability condition for the longitudinal grid cell size, given by $\Delta \xi^{\mathrm{c}} > 1/k_0$, $\Delta \xi$ should be chosen as close as possible to $\Delta \xi^{\mathrm{c}}$. On the other hand, since Eq.~(\ref{eqn:envelope}) must also resolve correctly the Rayleigh length, the spacing between successive s, $\Delta s$, must be chosen appropriately.

QuickPIC has been benchmarked with full 3D OSIRIS simulations for a number of situations, both for the plasma wakefield accelerator~\cite{bib:chengkun} and LWFA~\cite{bib:vieira}. Excellent agreement between the two codes was found. 

\section{One-to-one modeling of state-of-the-art experimental conditions}\label{sec:onetoone}

\subsection{Modeling of recent LWFA experiments}\label{subsec:exp}

We now examine the propagation of $\sim$ 2 J laser pulse using state-of-the-art experimental parameters~\cite{bib:leemans}, corresponding to the evolution of a $12~\mathrm{TW}$ and $25~\mathrm{TW}$ laser pulses, in both uniform and preformed parabolic plasma channels. We use $a_0=0.8$ and $a_0=1.1$ for the 12 TW and 25 TW laser pulses respectively. In both simulations, the laser interacts with a plasma with $n_0=3.4\times10^{18}\mathrm{cm}^{-3}$, and with $\Delta n=0.9\times10^{17}\mathrm{cm}^{-3}$ and $r_{\mathrm{c}}=110~\mu\mathrm{m}$, if the channel is present. The laser pulse duration is $\tau_0=73~\mathrm{fs}$, focused to $W_0=25~\mu\mathrm{m}$. The simulation box is $(360~\mu\mathrm{m})^2$ wide and $60~\mu\mathrm{m}$ long, divided into $512^2\times256$ cells for the transverse and longitudinal directions, respectively, with $4$ particles per cell.

\begin{figure}[htbp]
\begin{center}
\includegraphics[width=\columnwidth]{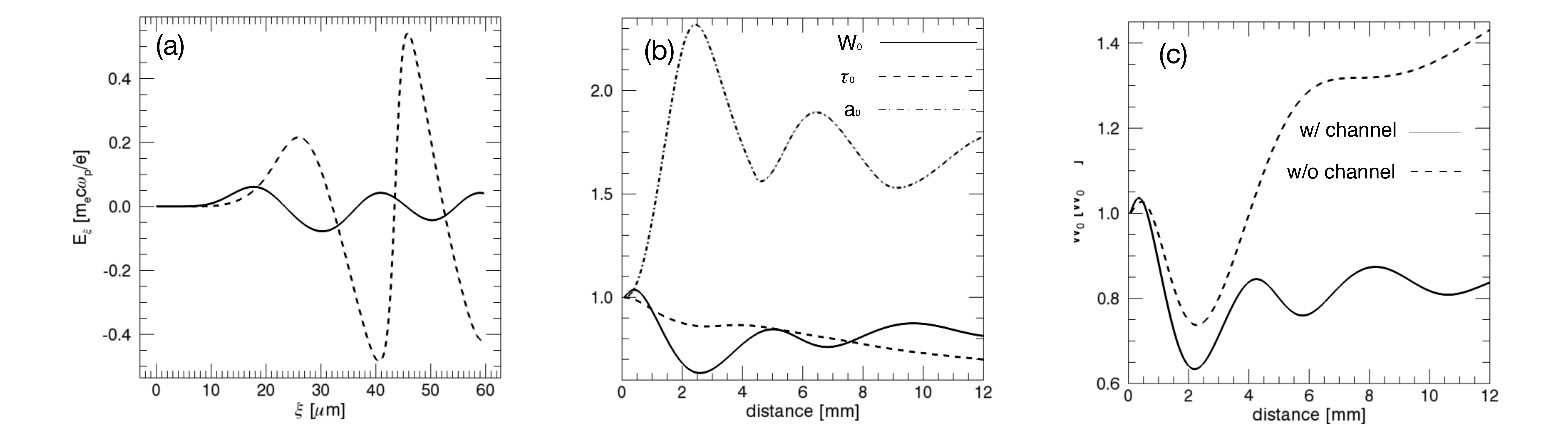}
\caption{\label{fig:12tw} (a) shows the accelerating fields at initially at $s=0$ (solid) and at $s=2.5~\mathrm{mm}$ (dashed), after the initial self-focusing stage and (b) shows the laser spot size, $W_0$, length, $\tau_0$, and peak $a_0$ evolution, normalized to the initial values, for the 12 TW laser simulation. Both (a) and (b) refer to the situation where the plasma channel is present. (c) shows the comparison between the evolution of the spot of the laser with (solid) and without (dashed) preformed plasma channel.}  
\end{center}
\end{figure}

We first examine the laser evolution of the "lower" power pulse in the plasma channel. Initially, the laser slightly disturbs the plasma, creating a perfectly linear wake with an amplitude well below the threshold for wave-breaking and self-injection (Fig.~\ref{fig:12tw}-a). After entering the plasma, the pulse is strongly self-focused leading to a fast compression of the spot-size to roughly $15~\mu\mathrm{m}$ after only $2.5~\mathrm{mm}$ of propagation. Simultaneously, the peak vector potential of the laser increased to $a_0\sim2$ (Fig.~\ref{fig:12tw}-b) leading to an increase of the peak power of the pulse by nearly a factor of 2. In addition to self-focusing, this also resulted from longitudinal self-modulation effects~\cite{bib:tsungpnas}. As the laser is modulated and the transverse and longitudinal ponderomotive forces become stronger, electrons are progressively expelled from the axis, leading to nearly full electron cavitation at $s\approx2~\mathrm{mm}$. By then, the accelerating gradient attains its maximum value, as shown in Fig.~\ref{fig:12tw}-a, suggesting that if self-injection occurs, it would be held most likely near that region. 


For $s>2~\mathrm{mm}$, the laser remains guided with spot oscillations of  roughly $20\%$ with a wavelength $\Delta s=2~\mathrm{mm}$. In order to understand if the guiding of the laser is only due to self-focusing, or due to the presence of the channel, we have performed a similar simulation in an uniform plasma. The simulation reveals that for $s>2~\mathrm{mm}$ the laser continuously diffracts, thus making the channel of fundamental importance for the guiding of the laser.

\begin{figure}[htbp]
\begin{center}
\includegraphics[width=\columnwidth]{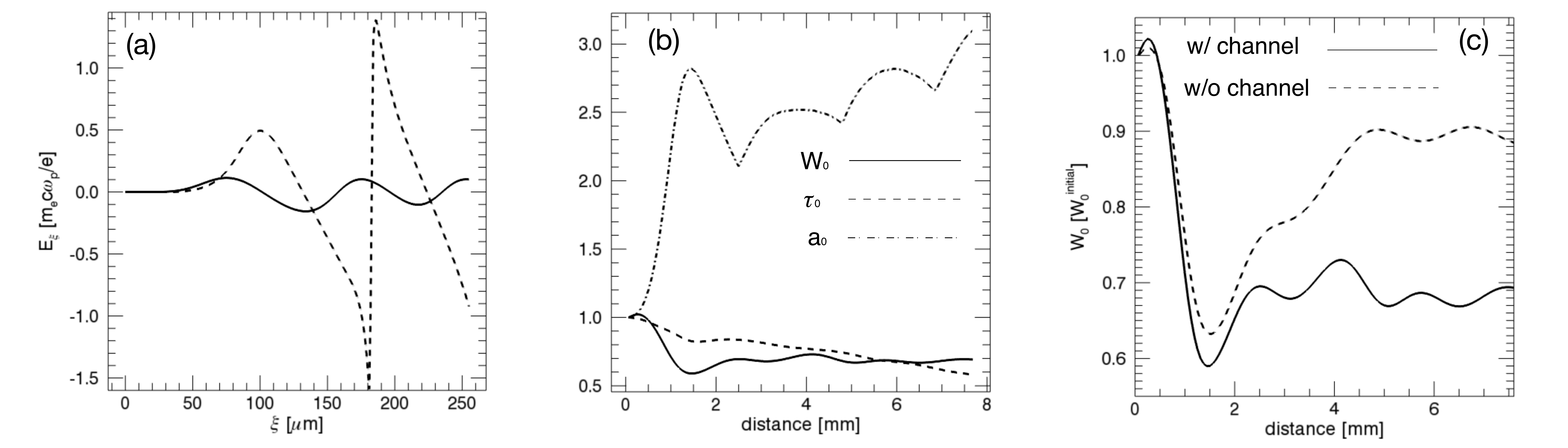}
\caption{\label{fig:25tw} (a) shows the accelerating fields at initially at $s=0$ (solid) and at $s=1.5~\mathrm{mm}$ (dashed), after the initial self-focusing stage. (b) shows the laser spot size, $W_0$, length, $\tau_0$, and peak $a_0$ evolution, normalized to the initial values for the 25 TW laser simulation. Both (a) and (b) refer to the simulations in the presence of the plasma channel. (c) shows the comparison between the evolution of the spot of the laser with (solid) and without (dashed) preformed plasma channel.}  
\end{center}
\end{figure}

The main features associated with the 12 TW laser pulse propagation in the channel are also present in the 25 TW laser simulation, also in a channel, as seen in Fig.~\ref{fig:25tw}. Despite being more intense, the initial wake is still below the threshold for wave-breaking and self-injection. However, as the laser propagates, the wake becomes stronger since both the laser spot is compressed due to self-focusing, and the peak vector potential is increased due to a combination of self-focusing, self-steepening and self-compression (Fig.~\ref{fig:25tw}-a). For $s=1.5~\mathrm{mm}$, the spot is compressed to its minimum value of $14~\mu\mathrm{m}$, and nearly all electrons are expelled from the laser axis. This leads to the excitation of a strongly non-linear wake (Fig.~\ref{fig:25tw}-a), where the occurrence of wave-breaking and self-injection becomes likely. 

A distinct behavior, in comparison to the "lower" power laser scenario, occurs for $s>1.5~\mathrm{mm}$, where the peak vector potential keeps rising (Fig.~\ref{fig:25tw}-b). This behavior is associated with the higher initial $a_0$ of the 25 TW laser, which further enhanced longitudinal self-modulations. In order to understand the role of the channel in this mechanism, an identical simulation in an uniform plasma was performed, which revealed that the laser pulse remained guided even for $s>1.5~\mathrm{mm}$ (Fig.~(\ref{fig:25tw}-c)). Thus, both longitudinal self-modulations, in particular self-compression which increases the peak laser vector potential, and transverse modulations which are further enhanced by the latter, are sufficient to reach stable laser propagation. We note, however, that for propagation distances larger than those we have modeled ($s_{\mathrm{max}}\simeq 1~\mathrm{cm}$), the laser may not be self-guided as the transverse profile of the laser may develop higher order gaussian beams. Furthermore, it is important to observe that in the 25 TW scenario the self-guiding is directly associated with the structure of the bubble \cite{bib:lu,bib:lu2}.

\subsection{Multi-GeV electron acceleration in the blowout regime}

Until now, the LWFA experimental results have been obtained in the self-injection regime. However, the external injection of an electron beam to be accelerated by a laser driven plasma wave, constitutes a rather promising path to produce quasi-monoenergetic, multi-GeV electron beams, without resorting to self-injection, because both the quality of the injected beam and the optimal accelerating properties of the blowout region can be carefully controlled. In this Section we then examine the dynamics of an externally injected beam, loaded in the bubble, driven by a $10~\mathrm{J}$ and $15~\mathrm{J}$ laser pulse. 

The externally injected electron beam is defined by a bi-gaussian distribution:

\begin{equation}
\label{eqn:beam}
n_b=n_{b0} \exp({-r^2/2 \sigma_r^2})\exp(-\xi^2/2 \sigma_{\xi}^2),
\end{equation}
where $n_{b0}$ is the peak electron beam density, $\sigma_r$ and $\sigma_{\xi}$ are the transverse width and duration of the beam. We note that QuickPIC treats self-consistently the fields associated with the external electron beam.
 
We start by examining the propagation of a 10 J laser pulse, following  \cite{bib:mangles2}, in an uniform plasma with $n_0=10^{18}~\mathrm{cm}^{-3}$. The laser is focused to $W_0=34~\mu \mathrm{m}$, has a duration of  $\tau_0=30~\mathrm{fs}$ corresponding to a peak vector potential $a_0=3$. The beam is placed close to the rear part of the first plasma wave, in the region of the highest accelerating fields. The total charge of the beam is $11.2~\mathrm{pC}$, with $\sigma_r=\sigma_{\xi}=1.5~\mu\mathrm{m}$, roughly corresponding to the dimensions of a self-injected bunch in similar conditions. The beam charge was chosen to be far from the beam loading limit. The beam was injected with $\gamma=40$, higher than the relativistic factor associated with the plasma wave phase velocity, which guarantees that the beam is trapped from the beginning of the simulation. The laser enters a simulation box with $(330~\mu\mathrm{m})^2$ wide and $60~\mu\mathrm{m} $ long, divided into $128^3$ cells, with 4 particles per cell. 

In contrast with the simulations presented in Sec.~\ref{subsec:exp}, plasma electrons are almost fully blown-out from the laser axis right at the entrance of the plasma, creating a plasma channel which self-guides the laser with spot variations on the order of $30\%$. These transverse oscillations, however, do not interfere with the acceleration process, as beam electrons are accelerated with a constant acceleration gradient of roughly $\approx 0.85~\mathrm{GeV/cm}$. 

An important property of the electron beam is the energy spread, which should be as small as possible. Since the electric field is stronger near the back of the bubble, beam electrons which are closer to that location accelerate faster. Naturally, this leads to an energy chirp and to the variation of the energy spread of the beam, which grows from $\Delta E^{\mathrm{initial}}/E<1\%$ to $\Delta E^{\mathrm{max}}/E \approx 20 \%$, at FWHM, until $s\lesssim1~\mathrm{cm}$. 

\begin{figure}[htbp]
\begin{center}
\includegraphics[width=\columnwidth]{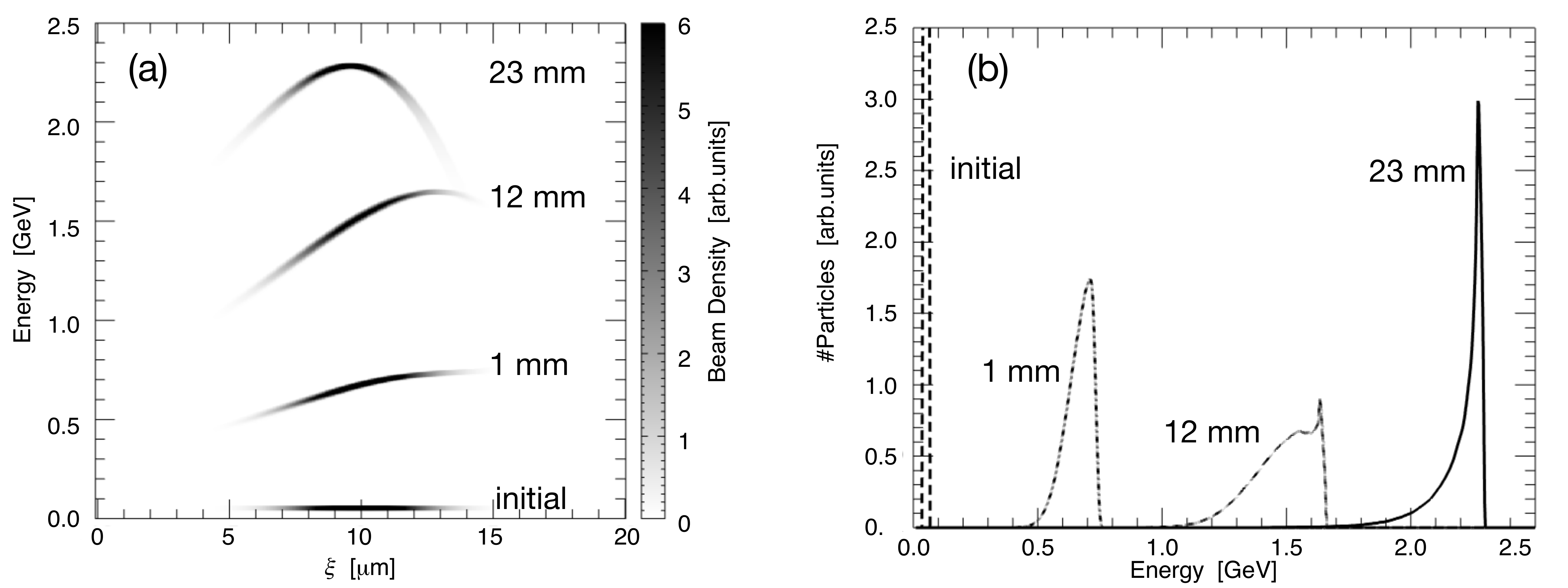}
\caption{\label{fig:10j} (a) shows the evolution of the beam phase space for the 10 J laser simulation, where the phase rotation of the beam occurred for $s=2.3~\mathrm{cm}$. (b) shows the corresponding spectra revealing a quasi-mono-energetic electron beam with $\approx 4\%$ energy spread and with $\sim 2.2~\mathrm{GeV}$ after phase space rotation occurred.}  
\end{center}
\end{figure}

For $s>1~\mathrm{cm}$ the opposite effect occurs. As the laser becomes pump depleted, the plasma wavelength decreases in such way that the front of the beam accelerates faster than the back. This effect progressively balances the beam energy chirp acquired until $s\lesssim1~\mathrm{cm}$, while it reduces its energy spread (phase-space rotation) thus leading to the formation of a quasi-monoenergetic bunch. After $s=2.3~\mathrm{cm}$, the beam then becomes nearly mono-energetic, with $\Delta{ E}^{\mathrm{final}}/E \approx 4 \%$ and with $E\approx 2.2~\mathrm{GeV}$, as shown in Figure~\ref{fig:10j}. We note that this phase-space rotation mechanism is different from that of Ref.~\cite{bib:tsung}. In that case, the generation of the quasi-mono-energetic electrons was due to the dephasing of the beam in the plasma wave while in the present situation, the phase rotation of the beam is due to the pump depletion of the laser. 

It is possible to obtain an extremely stable acceleration regime, with negligible laser and wake oscillations, by using a preformed plasma channel and by carefully choosing the remaining laser parameters. We have examined the acceleration of an externally injected beam in the wake of a 15 J laser pulse, with the laser and plasma parameters chosen according to the scaling laws to the blowout regime~\cite{bib:lu,bib:lu2}. The most stable laser propagation is obtained if the laser spot size is matched to the blowout radius of the electrons, $r_b\approx 2\sqrt{a_0} c/\omega_p$, which leads to $W_0\simeq r_b\simeq 37 \mu \mathrm{m}$, for $a_0=2$. In order to optimize the energy transfer from the laser to the beam electrons, the pump depletion length is matched to the dephasing length, which leads to $\tau_0=87~\mathrm{fs}$. The above considerations also imply that the background density is given by $n_0=1.6\times10^{17}\mathrm{cm}^{-3}$. We use a channel depth $\Delta n= 6\times10^{16}\mathrm{cm}^{-3}$, which is $3/4$ of the linear guiding theory matching condition for a parabolic channel. We use an externally injected electron beam with $11.2~\mathrm{pC}$, chosen to be far from the beam loading limit, placed in the region of maximum accelerating fields, with $\sigma_r=\sigma_{\xi}=3~\mu\mathrm{m}$,  roughly corresponding to the dimensions of a self-injected bunch in similar conditions. The beam is injected with $\gamma=200$, which is higher than the relativistic factor associated with the wave phase velocity, as determined from \cite{bib:lu2,bib:decker} guaranteeing that particles are immediately trapped at the beginning of the propagation. The laser enters the simulation box with $(612~\mu\mathrm{m})^2$ wide, $140~\mu\mathrm{m} $ long, divided into $128^2\times 256$ cells for the transverse and longitudinal directions respectively, with 4 particles per cell.
 
These parameters provide a nearly non-evolving acceleration structure, where the laser spot size oscillates within only $3 \%$ of the initial value. As the laser enters the plasma, it creates a mildly nonlinear wakefield, where the beam electrons dephase in the presence of an accelerating gradient of $200~\mathrm{MeV}/\mathrm{cm}$, for the peak of the energy distribution. This is in good agreement with the theoretical predictions~\cite{bib:lu,bib:lu2} which predict an accelerating gradient of $260~\mathrm{MeV}/\mathrm{cm}$. At the end of the simulation, after 20 cm of propagation, beam electrons gained more than 4 GeV. The theoretical maximum energy gain is $\approx 7~\mathrm{GeV}$ after $26~\mathrm{cm}$, but for such long distances the laser envelope modulations are too strong, and the envelope approximation, critical for the laser field evolution in QuickPIC, breaks-down due to strong self-compression and self-steepening. 

\section{Discussion and conclusions}\label{sec:conclusions}

Since QuickPIC simulations for such long plasmas are already quite large, and for some scenarios, kinetic effects not modeled by QuickPIC, can be relevant, we have assessed the possibility to use full 2D PIC simulations for quantitative predictions in long propagations distances. We have compared the results presented in Sec.~\ref{sec:onetoone} with full 2D simulations in OSIRIS, using the same physical parameters for the laser and plasma channel. For the 2D OSIRIS simulation, the computational window, that moves at the speed of light, is $230~\mu\mathrm{m}$ wide and $60~\mu\mathrm{m}$ long. The box is divided in $400\times2010$ cells for the transverse and longitudinal directions, respectively, with 9 particles per cell.

\begin{figure}[htbp]
\begin{center}
\includegraphics[width=\columnwidth]{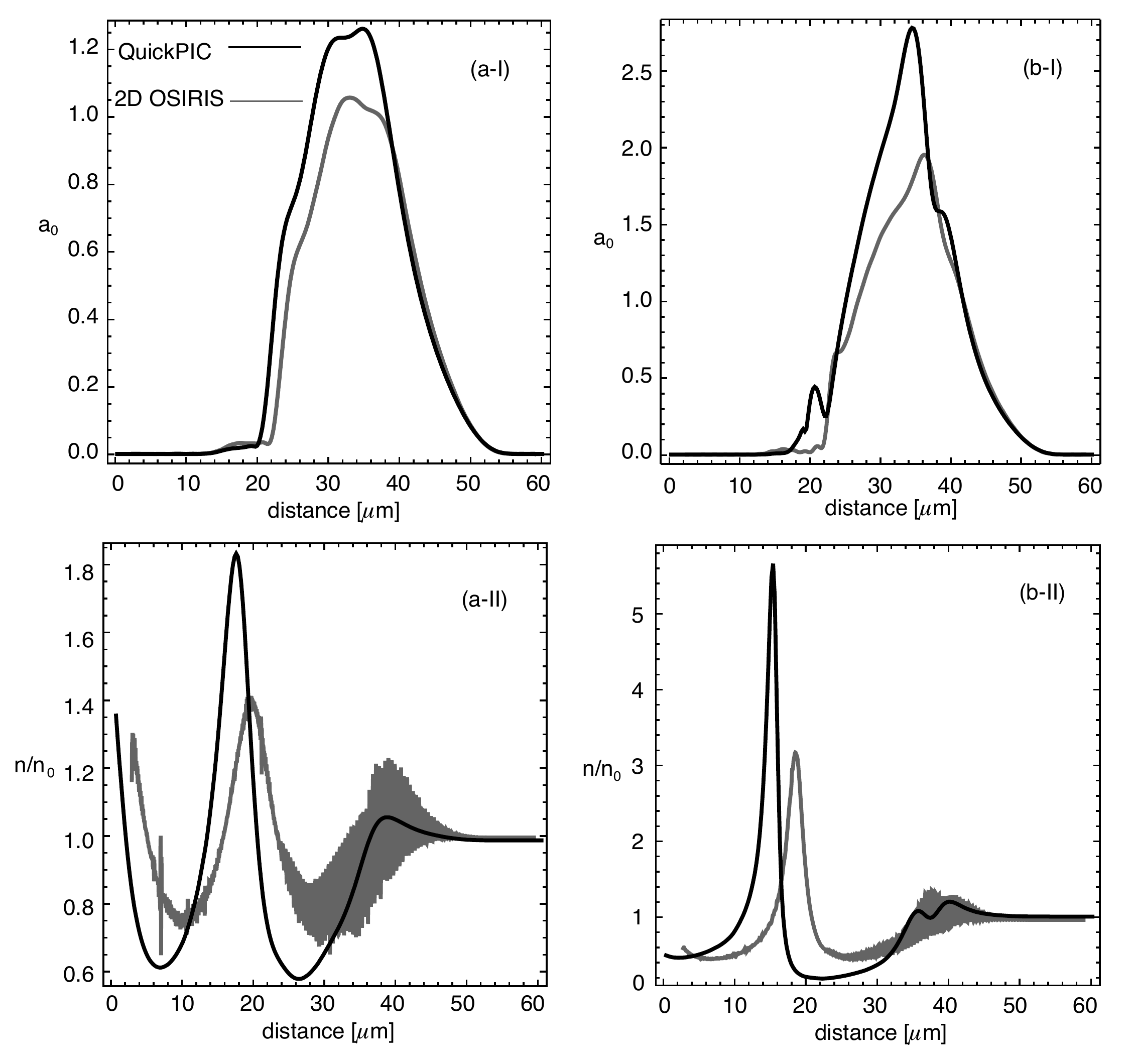}
\caption{\label{fig:2dosiris} Comparisons between the 2D OSIRIS and QuickPIC simulations after $s=3.75~\mathrm{mm}$. (a-I) shows the on-axis longitudinal laser profile and (a-II) the on-axis electron density distribution for the $12 \mathrm{TW}$ laser pulse. (b-I) shows the on-axis longitudinal laser profile and (b-II) the on-axis electron density distribution for the $25~\mathrm{TW}$ laser.}  
\end{center}
\end{figure}

The comparison between QuickPIC and the 2D simulations is shown in Fig.~\ref{fig:2dosiris}. The lower dimensionality of the 2D simulations lead to quantitative differences, in comparison with QuickPIC. In the 2D full PIC simulation, the laser evolution is weaker, even though the threshold power for self-focusing, given by $P/P_c=a_0^2 W_0^2/16\sqrt{2}$, is lower than in 3D. This leads to differences of $\approx 30\%$ on the laser peak $a_0$. Naturally, the wakefield becomes weaker and shorter for the 2D OSIRIS simulations and was not even strong enough to self-inject plasma electrons~\cite{bib:tsung}. Even so, and despite the quantitative differences, 2D PIC simulations and QuickPIC are qualitatively similar. 

These results then indicate that the numerical modeling of LWFA experiments, can be performed accurately only in 3D geometries \cite{bib:puck2}. However, the use of standard full PIC simulations becomes prohibitive, as the computing requirements increase with the interaction length associated with higher energy gains. Therefore, the use of QuickPIC or full PIC 3D simulations in boosted frames~\cite{bib:vay} (currently being implemented in OSIRIS~\cite{bib:martins,bib:fonseca2}) are valuable tools to examine LWFA under these conditions. 

The use of QuickPIC has allowed us to perform full-scale, one-to-one numerical experiments of LWFA using state-of-the-art experimental parameters in scenarios where full PIC 3D simulations are difficult to perform. We have examined the laser propagation in conditions which closely follow recent experiments \cite{bib:leemans}; for $\sim 10~\mathrm{J}$ laser pulses multi-GeV electron acceleration has been observed. 

We have shown that recent experiments have been dominated by a strong initial self-focusing. This stage is then followed by small spot size oscillations. In the case of the 12 TW laser, the presence of the channel is crucial in order to guide the pulse after the strong initial self-focusing. In the 25 TW laser scenario, the channel only confines the laser to smaller spot sizes as the laser propagates. Unlike the 12 TW laser, the propagation of the 25 TW laser in the pre-formed plasma channel, lead, after the initial self-focusing stage, to the continuous increase of the laser peak vector potential. The higher initial $a_0$ for the 25 TW laser then resulted in the enhancement of both longitudinal and transverse self-modulations, which drove a strong enough blowout region to guarantee guiding even in uniform plasma.

For 10 J laser pulses, we have shown that it is possible to accelerate electrons up to 2.2 GeV in 1.5 cm, in an uniform plasma. By using a 15 J laser pulse, it is possible to generate at least 4 GeV electrons in 20 cm long plasma channels. The use of a smaller $a_0$ in this latter scenario in comparison to the simulation of the 10 J laser pulse, implied the use of a smaller background density in order to guarantee the matching conditions proposed by W. Lu et al \cite{bib:lu2}. Since the dephasing length is higher for lower densities, the maximum energy gain is higher in this situation. We have shown that the electron beam can be efficiently accelerated in long plasmas, thus indicating that in the near future, with the systems now coming online, mono-energetic electron beams can be accelerated to multi-GeV energies.

\section*{acknowledgments}
Work partially supported by FCT (Portugal) through the grants PTDC/FIS/66823/2006 and SFRH/BD/22059/2005, and by the European Comunity - New and Emerging Science and Technology Activity under the FP6 "Structuring the European Research Area" programme (project EuroLeap, contract number 028514). The simulations were performed at the expp and IST clusters in Lisbon.

\begin{figure}[htbp]
\begin{center}
\includegraphics[width=\columnwidth]{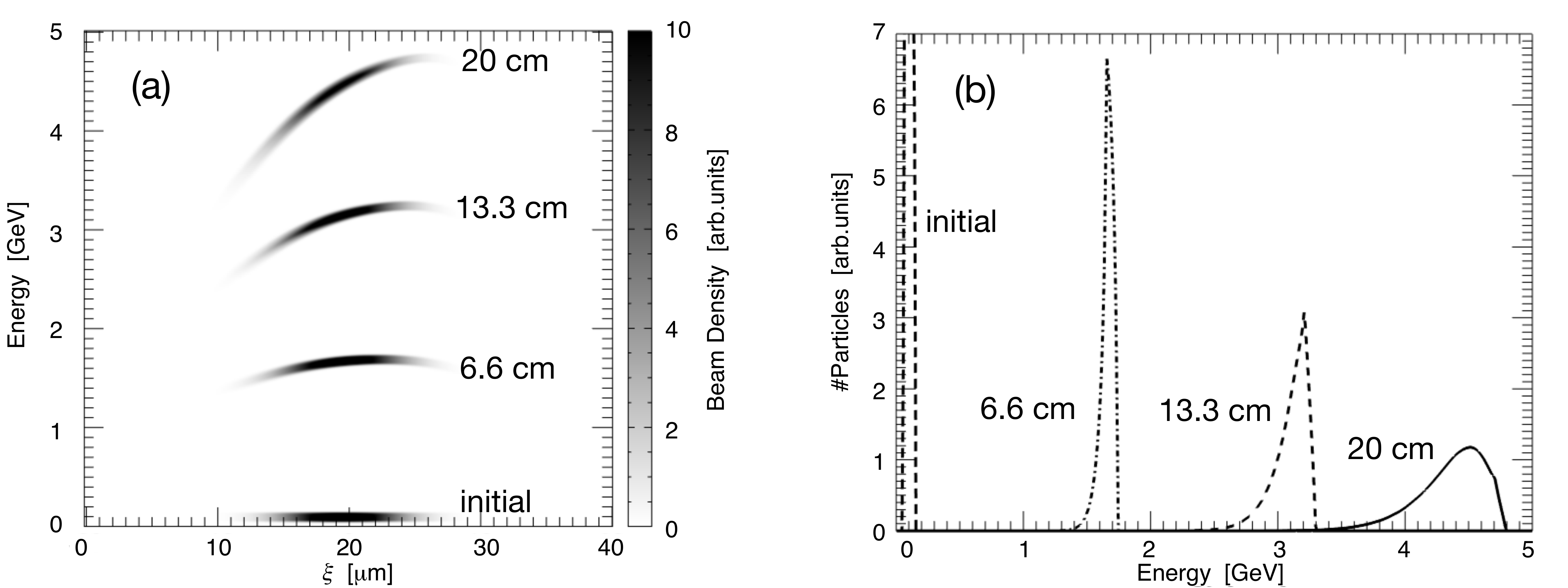}
\caption{\label{fig:15j} (a) shows the evolution of the beam phase space for the 15 J laser simulation with external guiding. (b) shows the corresponding energy spectra, where the peak of the beam accelerated to $\approx4.2~\mathrm{GeV}$}  
\end{center}
\end{figure}

\end{document}